\documentclass[journal,twoside]{IEEEtran}
\usepackage{cite}
\usepackage{amsmath,amssymb,amsfonts}
\usepackage{algorithmic}
\usepackage{graphicx}
\usepackage{textcomp}
\usepackage{subcaption}
\usepackage{graphicx}

\usepackage{multirow}
\usepackage{mathtools}
\usepackage{nccmath}
\usepackage[section]{placeins}
\usepackage{graphicx}
\usepackage{tikz}
\usepackage{subcaption}
\usepackage{booktabs}
\usepackage{mathptmx}

\mathchardef\mhyphen="2D
\def\BibTeX{{\rm B\kern-.05em{\sc i\kern-.025em b}\kern-.08em
    T\kern-.1667em\lower.7ex\hbox{E}\kern-.125emX}}
\begin{document}

\title{Power Flow Solution in Unbalanced 3-Wire MV and 4-Wire LV Networks Using Symmetrical and Eigen-basis Coordinates}
\author{First A. Author, \IEEEmembership{Fellow, IEEE}, Second B. Author, and Third C. Author, Jr., \IEEEmembership{Member, IEEE}
\author{Abduljalil S. Aljadani, Firdous U. Nazir, \IEEEmembership{Senior Member, IEEE}, 
    Bikash C. Pal,\IEEEmembership{Fellow, IEEE},\\
   Izudin Džafić, \IEEEmembership{Senior Member, IEEE}, and Rabih A. Jabr, \IEEEmembership{Fellow, IEEE}.}}
\maketitle

\begin{abstract}
The large penetration of distributed generations impacts both the secondary low-voltage (LV) and the primary medium-voltage (MV) segments of the distribution network. Optimizing power flow calculations for the integrated MV/LV networks is crucial for the real-time management of modern distribution networks. Traditional methods in symmetrical coordinates are primarily limited to the three-wire model of three-phase networks, often leading to inaccuracies in power flow calculations when applied to three-phase four-wire LV segments. This paper introduces a novel power flow method for integrated three-wire MV and four-wire LV networks. Using eigenvector decomposition to diagonalize the admittance matrix of four-wire LV lines, the proposed method improves the computational efficiency of power flow calculations and accurately calculates the neutral-to-ground voltage. The results of the case studies show over 50\% reduction in the number of non-zero elements in the LU factors of the bus admittance matrix, and speed-up factors of 2.78 on the IEEE 123-node test system and 3.63 on the IEEE 8500-node test system in execution times for Volt/Var control (VVC), compared to the phase coordinates model.
\end{abstract}


\section{Introduction}
\label{sec:introduction}
\IEEEPARstart{P}{ower} flow calculation is an essential step in the operation and control of power grids \cite{01}. Routine tasks such as Volt/Var control (VVC) and optimal feeder reconfiguration often require several power flow runs. Following the requirements of real time management of the system, it is important to use power flow methods capable of providing stable and accurate solutions within minimal computational time \cite{11}. LV distribution networks are characterized by unsymmetrical three-phase four-wire configuration with a grounded neutral. The combination of this unsymmetrical configuration and unbalanced loading leads to unequal voltage magnitudes between phases, resulting in the flow of currents in the neutral wires leading to the presence of the neutral-to-ground voltage. The consideration of the neutral-to-ground voltage is important to account for multiple solutions in power flow calculation and to accurately calculate the system transfer capacity \cite{impacts}, \cite{Opt}.
\subsection{Related work}
Detailed models of LV distribution lines and groundings are required to accurately calculate the power flow. In \cite{linemodels}, a phase coordinates model for single-, two-, and three-phase laterals with a neutral wire was presented. In this model, a modified form of Carson’s equations was used to compute the self and mutual impedances of line segments, enabling accurate representation of LV distribution feeders. This model was used in \cite{Penido} to calculate the power flow in three-phase four-wire distribution networks. This work extended the three-wire current injection method (TCIM) to a four-wire formulation (FCIM), allowing direct calculation of neutral voltages and currents using Newton–Raphson. The TCIM and FCIM were applied in \cite{ALAM} to solve the power flow in integrated three-wire MV and four-wire LV distribution networks. In \cite{SUNDERLAND201630}, an approach was proposed to model generators and loads as constant shunt admittances connected in parallel to a correction current injections. Current injections were also used to model the shunt admittances of lines and transformers \cite{IETGeneration}. Using current injections reduced the number of iterations and improved convergence stability compared to Newton and forward–backward sweep methods. In \cite{Ciric}, the four-wire line model was extended to a five-wire model, including the ground return path. The ground was modeled as a perfect conductor, and 5x5 line impedance matrix was calculated using Carson’s equations. This model enabled the analysis of distribution networks under open neutral conditions \cite{openneutral}.

The application of the Fortescue transformation resulted in significant improvements in computation time and memory requirements of power flow algorithms in large practical networks \cite{2}.  In \cite{3}, a symmetrical coordinates Bus admittance power flow method (Y\textsubscript{BUS}) for three-phase three-wire networks has been proposed. The mutual coupling of untransposed lines has been modeled as current injections; decoupling the positive, negative and zero sequence networks. The use of current injections eliminated the approximation errors introduced by the symmetry assumptions on untransposed lines \cite{linemodels}. As a result, the size of the Y\textsubscript{BUS} matrix is reduced for an N-Bus network from (3N x 3N) to (N x N) when compared with the phase coordinates model. In \cite{FAST}, \cite{mam1}, \cite{mm1} and \cite{mm2} the decoupled positive sequence network has been solved using the Newton Raphson (NR) and fast decoupled methods. These methods reduced the computation time for systems with distributed generation (DG) in comparison to the Y\textsubscript{BUS} method. Also, symmetrical coordinates models of transformers, DGs and voltage source converters (VSC) interfaced distributed energy resources have been used.  

In \cite{mam2}, a power flow method for solving distribution networks with single, two and three phase laterals has been proposed. In this method the symmetrical three-phase segment of the network was solved using NR in symmetrical coordinates, the single and two phase segments were solved in phase coordinates using backward/forward sweep. To solve the power flow, an iterative process was used where the single and two phase segments of the network were solved first to calculate current injections to the three phase segment. The three phase segment was then solved to calculate the bus voltages of the non-three phase segments. In \cite{14}, a method to model  single, two and three phase laterals in symmetrical coordinates has been proposed. In this method, Fortescue transformations of order one and two have been used to transform the admittance matrices of single and two phase laterals to symmetrical coordinates. These models enabled solving the power flow using backward/forward sweep in symmetrical coordinates, eliminating the need for solving single and two phase segments in phase coordinates. In \cite{15}, the symmetrical coordinates models in \cite{14} were adopted to build the network bus admittance matrix. As a result, power flow in networks with non three-phase segments can be calculated using the Y\textsubscript{BUS}, NR and fast decoupled methods.

The use of symmetrical coordinates in power flow calculation is currently limited to the three-wire model of three phase networks \cite{Opt}, \cite{low}. In this model, the neutral node is eliminated using Kron's reduction, which assumes zero voltage drop across the neutral. Kron's reduction is normally used in modelling MV segments of distribution networks \cite{4}. LV segments are unsymmetrical with unbalanced loads, and the neutral wire is grounded by resistances at regular intervals (multi-grounded). A Kron-reduced model eliminates the neutral node of the LV segments, which neglects the impact of the neutral-to-ground voltage in power flow calculations and leads to inaccuracies in calculating the network voltages and currents \cite{Opt}. To overcome the limitations of the symmetrical coordinates model, in this paper eigenvector decomposition is proposed to diagonalize the admittance matrix of LV segments and transform voltages and currents to eigen-basis coordinates.

\subsection{Technical contribution}
This paper proposes a model of integrated three-wire MV and four-wire LV networks in symmetrical and eigen-basis coordinates and contributes the following:\begin{itemize}
    \item A method to diagonalize the line admittance matrix of four-wire LV networks. This method uses eigenvector decomposition to compute the eigenvalues and eigenvectors of the line admittance matrix. The eigenvalues are used to construct the diagonalized admittance matrix, whereas the eigenvectors form the transformation matrix that maps currents and voltages into eigen-basis coordinates.
 \item Sensitivity analysis of the parameters of Carson’s equations. This analysis shows that within the practical range of conductors and pole configurations, the LV feeder line admittance matrix is diagonalizable.
 \item Technical details of transforming the Bus admittance matrix of integrated MV/LV distribution networks from phase coordinates to symmetrical and eigen-basis coordinates, and decomposing the integrated MV/LV distribution network to smaller sub networks that can be solved in parallel.
 \end{itemize} 
 Power flow calculation is the most computationally intensive part of the VVC procedure \cite{ODCD}. The primary objective of VVC is to eliminate voltage violations across the network, with a secondary objective of minimizing network losses \cite{VVC}. This dual objective is formulated as an optimization problem that aims to minimize network losses while maintaining voltage magnitudes and branch current magnitudes within their prescribed limits. The Discrete Coordinate-Descent (DCD) algorithm is a standard approach used in industry for practical VVC implementations \cite{sensDCD}. This algorithm operates by iteratively updating one control variable at a time, thereby optimizing the system in discrete steps. A power flow calculation is needed at each optimization step, requiring a significant number of power flow calculations. Power flow calculations are made more efficient through the modeling of the network in symmetrical and eigen-basis coordinates, reducing both calculation time and memory requirement, and accurately calculating the neutral-to-ground voltage. 
\subsection{Paper structure}
The rest of this paper is organized as follows. Section II presents the proposed modelling of three-phase four-wire LV networks in eigen-basis coordinates. Section III presents a modeling example of a 4-bus integrated MV/LV feeder. This example illustrates the application of phase mapping between the different coordinates in the modeling of a MV/LV transformer. Section IV presents the decomposed Y\textsubscript{BUS} method, which is employed to solve the power flow in the proposed coordinates. Section V presents a series of case studies involving networks of varying sizes, that evaluate the accuracy, memory requirements of the proposed power flow method in comparison to the phase coordinates methods available in literature \cite{ALAM} and \cite{Das}. Section VI presents the sensitivity-based DCD algorithm and its application in VVC to evaluate the computational performance of the proposed power flow method.
\section{Modelling in Eigen-Basis Coordinates}
This section presents the proposed eigenvalue decomposition method to diagonalize the line admittance matrix, along with the sufficient conditions for diagonalization. This method can be applied to any line model (e.g. 5x5, 3x3, or 2x2) provided that the conditions outlined in Subsection \ref{oproof} are satisfied. The remainder of the paper focuses on three-phase, four-wire LV networks.
\subsection{Diagonalization using eigenvector decomposition}\label{diagss}
Consider a poly-phase network of $n$ phases. The network phasors can be expressed in terms of new $N$ vectors in a different set of coordinates using the following linear transformation: 
\begin{equation}\label{T1}
\begin{bmatrix}
V_a \\
V_b \\
V_c \\
\vdots \\
V_n
\end{bmatrix}
=
\begin{bmatrix}
a_{0,0} & a_{0,1} & a_{0,2} & \cdots & a_{0,n-1} \\
a_{1,0} & a_{1,1} & a_{1,2} & \cdots & a_{1,n-1} \\
a_{2,0} & a_{2,1} & a_{2,2} & \cdots & a_{2,n-1} \\
\vdots & \vdots & \vdots & \ddots & \vdots \\
a_{n-1, 1} & a_{n-1, 2} & a_{n-1, 3} & \cdots & a_{n-1,n-1} 
\end{bmatrix}
\begin{bmatrix}
V_0 \\
V_1 \\
V_2 \\
\vdots \\
V_N
\end{bmatrix}
\end{equation}
where the only restriction on the choice of coefficients in (\ref{T1}) is that the determinant of the transformation matrix is not zero. If the complex roots of unity (1,$a$,$a^2$....,$a^{n-1}$) are selected as coefficients, (\ref{T1})  will represent Fortescue transformation to symmetrical coordinates \cite{clarke}. For a known line admittance matrix $Y^{line}_{ph}$, the set of coefficients of the transformation matrix that diagonalizes $Y^{line}_{ph}$ can be computed using eigenvector decomposition as follows \cite{matrix}:
\begin{itemize}
    \item \textbf{Find the Eigenvalues}: Solve the characteristic equation (2) to find the eigenvalues \( \lambda_1, \lambda_2, \lambda_3, \lambda_4 \).
    \begin{equation}
    \det({Y^{line}_{ph}} - \lambda {I}) = 0 
    \end{equation}
    \item \textbf{Find the Eigenvectors}: For each eigenvalue \( \lambda_i \), solve (3) to find the corresponding eigenvector \( {v}_i \).
    \begin{equation}
    ({Y^{line}_{ph}} - \lambda_i {I}) {v}_i = 0
    \end{equation}
    \item \textbf{Form the Eigenvector Matrix}: Construct the matrix \({T_{eig}^{ph}} \) using the eigenvectors as columns:
    \begin{equation}
    {T_{eig}^{ph}}  = [{v}_1 \ {v}_2 \ {v}_3 \ {v}_4]
    \end{equation}
    \item \textbf{Diagonalize the Matrix}: The matrix \( {Y^{line}_{ph}} \) can be diagonalized as follows:
    \begin{equation}
    {Y^{line}_{ph}} =  {T_{eig}^{ph}} {Y^{line}_{eig}}  {T_{ph}^{eig}}
    \end{equation}
\end{itemize}
 where \( {Y^{line}_{eig}} \) is the diagonal matrix of the eigenvalues (\({Y^{line}_{eig}} = \text{diag}(\lambda_1, \lambda_2, \lambda_3, \lambda_4)\)), the matrix $T_{eig}^{ph}$ performs a transformation of the voltages and currents from eigen-basis to phase coordinates, the matrix $T_{ph}^{eig}$ performs the inverse transformation from phase to eigen-basis coordinates. Additionally, $T_{ph}^{eig}=(T_{eig}^{ph})^{-1}$.

\subsection{Diagonalizability of the LV feeder admittance matrix}\label{oproof}
\begin{figure}[t!]
    \centering
    \begin{subfigure}[b]{0.5\linewidth}
        \centering
        \includegraphics[ width=1.5 in]{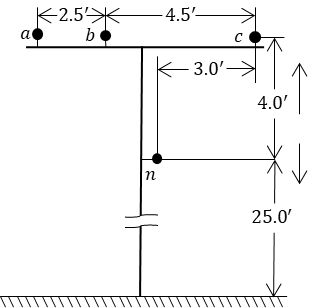}
        \caption{}
        \label{fig:sub1}
    \end{subfigure}
    \hfill
    \begin{subfigure}[b]{0.4\linewidth}
        \centering
        \includegraphics[ width=0.55 in]{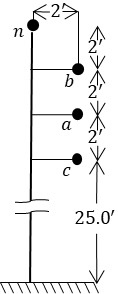}
        \caption{}
        \label{fig:sub2}
    \end{subfigure}
    \caption{Three-phase overhead line configurations: a) horizontal, b) vertical.}
    \label{fig:main}
    \label{pole}
\end{figure}

Now let us consider the configurations of the conductors for two common three-phase four-wire overhead lines shown in Fig. \ref{pole}. The line impedance matrix is normally computed using Carson’s equations \cite{kerst}:
\begin{equation}\label{Zii}
Z_{ii} = r_i + \pi^2 f G +j 4 \pi f G \left( \ln \frac{1}{GMR_i} + 7.6786 + \frac{1}{2} \ln \frac{\rho}{f} \right)
\end{equation}
\begin{equation}\label{Zij}
Z_{ij} = \pi^2 f G + j 4 \pi f G \left( \ln \frac{1}{D_{ij}} + 7.6786 + \frac{1}{2} \ln \frac{\rho}{f} \right)
\end{equation}
where \(r_i\) is the resistance per unit length of conductor \(i\), $GMR_i$ is the geometric mean radius, \(D_{ij}\) is the distance between conductors \(i\) and \(j\), \(\rho\) is the resistivity of the earth (assumed to be 100 $\Omega \mhyphen m$), $G$ is the resistance of the earth return path per unit length (normally assumed to be $0.1609\times 10^{-3}\hspace{0.1cm} \Omega$/mile \cite{kerst}), and \(f\) is the system frequency (assumed to be 60 Hz). 

From (\ref{Zii}) and (\ref{Zij}), it is evident that the impedance matrix is symmetric, and the diagonal elements are function of the conductor's characteristics. The real part of the off-diagonal elements ($R_{ij}$) is independent of the pole configuration and conductor's characteristics, the imaginary part ($X_{ij}$) is a function of the pole configuration. Table \ref{dist} shows the eigenvalues of the line impedance matrix of the horizontal and vertical pole configurations in Fig. \ref{pole} and the relative distance between the eigenvalues with respect to the eigenvalues of the impedance matrix of the horizontal configuration calculated using (\ref{Rdist}):
\begin{equation}\label{Rdist}
\text{Relative\hspace{0.1cm}Distance}= 100 \times \frac{|\lambda_H-\lambda_V|}{|\lambda_H|}
\end{equation}
 
\begin{figure}[t!] 
\centering
\includegraphics[ width=3.3 in]{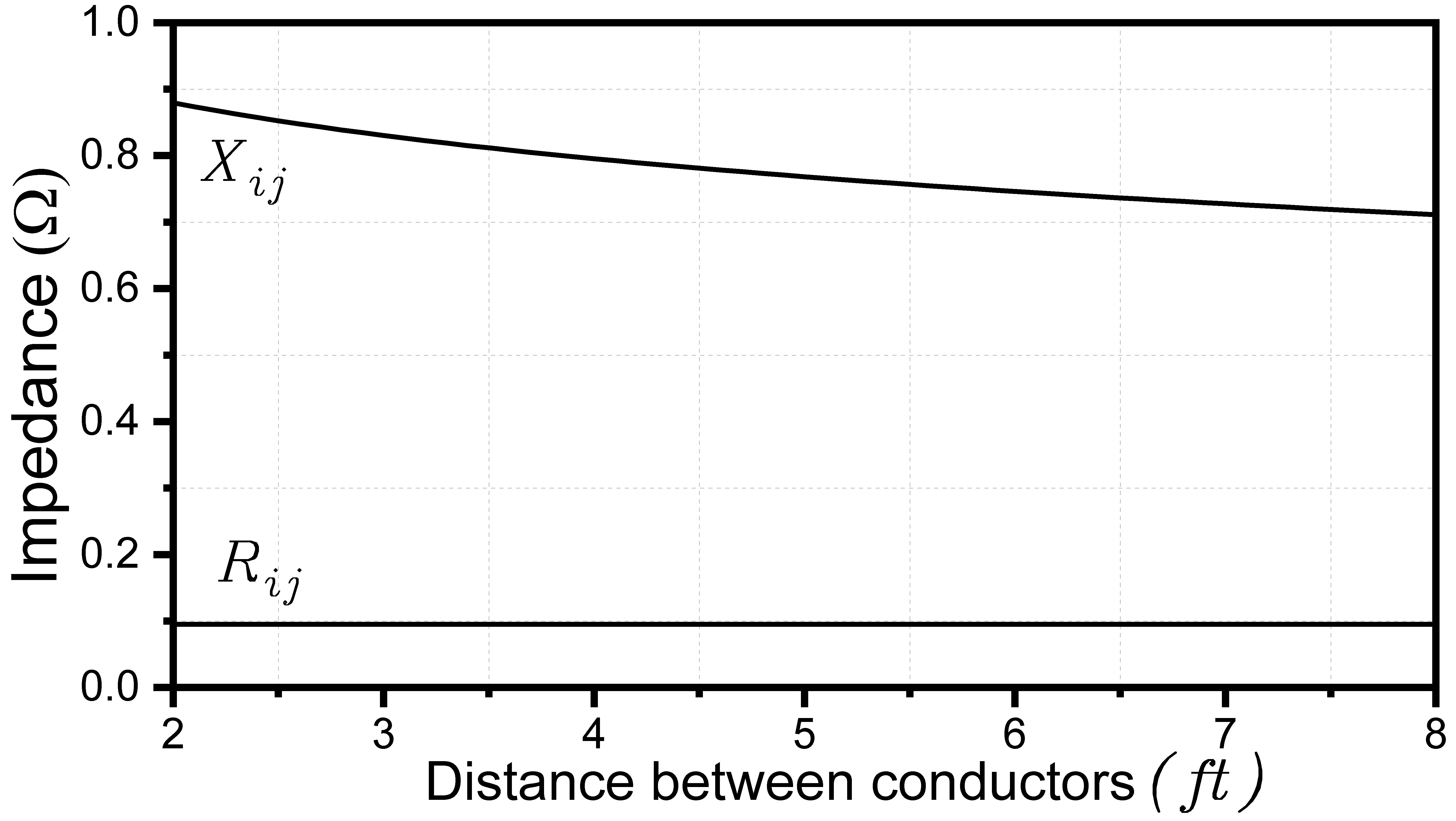}
\caption{Effect of the distance between conductors on $X_{ij}$, $R_{ij}$}
\label{distance}
\end{figure}

 The eigenvalues of the line impedance matrix of the two pole configurations fall within a close proximity with relative distances less than 12\%. This is because $X_{ij}$ is not significantly sensitive to changes in the pole configuration, as the term \(D_{ij}\) in (\ref{Zij}) is within a logarithmic function. Fig. \ref{distance} shows the values of $R_{ij}$ and $X_{ij}$ over a practical range of \(D_{ij}\) \cite{handbook}. The low sensitivity of the line impedance matrix to \(D_{ij}\) indicates that modifications to the pole configurations within practical limits will not alter the number of distinct eigenvalues, ensuring that the eigenvectors remain linearly independent. Consequently, the matrix is diagonalizable, as stated by the matrix diagonalization theorem \cite{handbook}:

 \textbf{ Matrix diagonalization theorem}: A square matrix $A$ of order $n \times n$ is diagonalizable if and only if there exists a set of $n$ linearly independent eigenvectors of $A$.

\begin{figure}[t!] 
\centering
\includegraphics[ width=3.3 in]{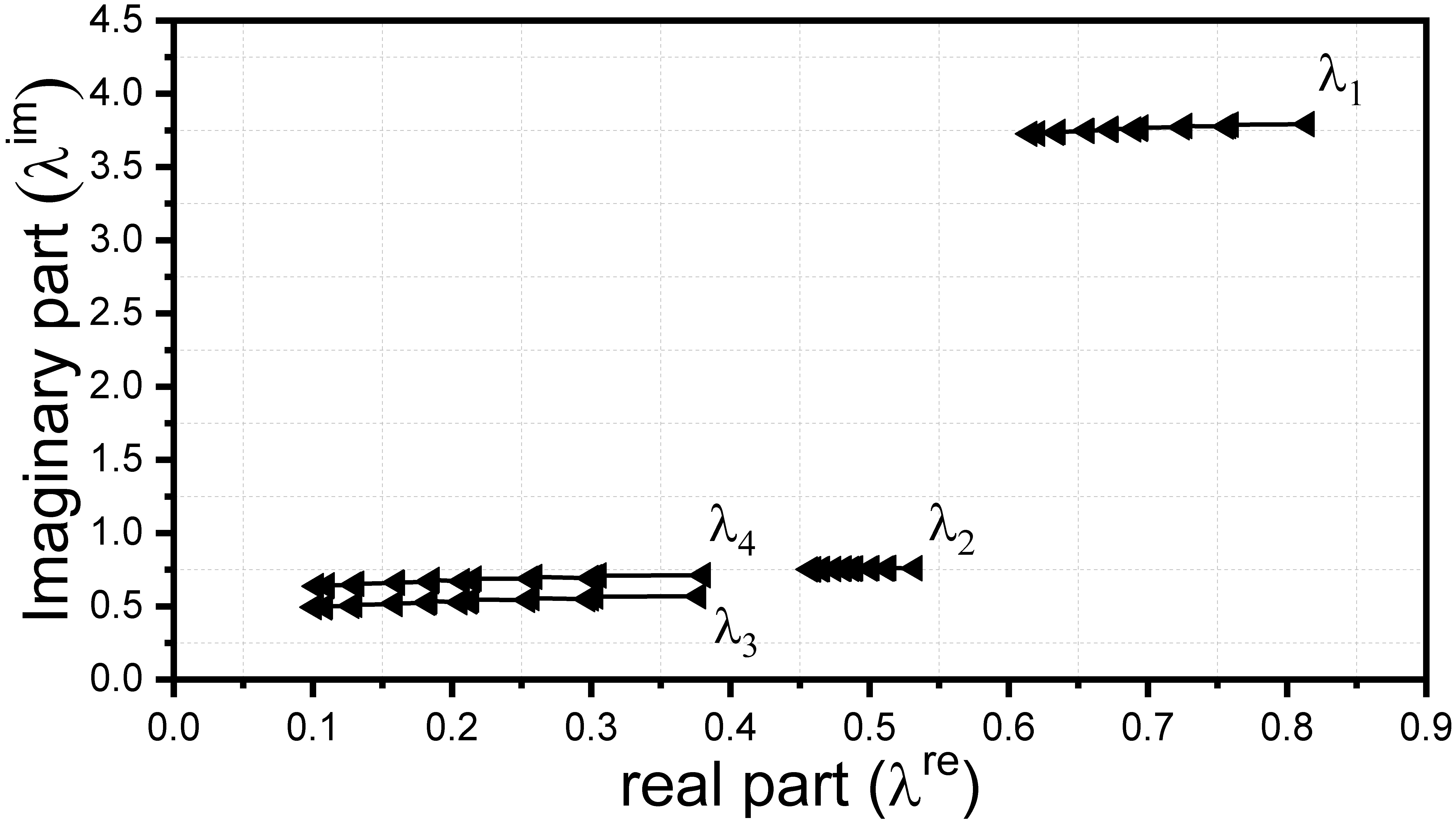}
\caption{Effect of the conductor size on eigenvalues of the line impedance matrix.}
\label{eigenvalues}
\end{figure}

Fig. \ref{eigenvalues} illustrates the eigenvalues of the line impedance matrix for the horizontal configuration over a wide range of phase conductor’s sizes (ACSR with diameters ranging from 0.642 to 1.213 in) \cite{handbook}. Increasing the conductor size decreases the real part of the eigenvalues, leading to an approximately horizontal shift. Consequently, within this practical range of conductors, the line impedance matrix remains diagonalizable. 

\section{A Modelling Example}\label{bui}
Consider the phase coordinates model of the 4-bus integrated MV/LV distribution feeder shown in Fig. \ref{fig:MV_LV}. This feeder consists of a three-wire MV segment, a delta-wye step-down transformer between Bus 2 and Bus 3, and an unsymmetric four-wire LV segment with its neutral wire grounded through resistances. The subsequent sections will provide a detailed modeling of the network components.
\
\begin{table}[t!]
  \centering
  \caption{\textsc{ Eigenvalues of The Line Impedance Matrix of Different Pole Configuration}}
  \label{dist}
    \renewcommand{\arraystretch}{1.1}
  \resizebox{0.95\linewidth}{!}{%
 \fontsize{10}{10}\selectfont 
    \begin{tabular}{cccc}
      \toprule
 \multirow {2}{*}{Eigenvalue} & \multirow {1}{*}{Horizontal} & \multirow {1}{*}{Vertical} & \multirow {1}{*}{Euclidean Distance}  \\
                            &     {Configuration}            &    {Configuration}        &    {in Complex Plane}  \\
\midrule
\(\lambda_1\) & \(0.7624 + 3.7800i\) & \(0.7591 + 3.9008i\) & \(3.13 \%\) \\
\(\lambda_2\) & \(0.5141 + 0.7582i\) & \(0.5076 + 0.7747i\) & \(1.93\%\) \\
\(\lambda_3\) & \(0.3061 + 0.5543i\) & \(0.3072 + 0.5048i\) & \(7.81\%\) \\
\(\lambda_4\) & \(0.3085 + 0.6982i\) & \(0.3172 + 0.6103i\) & \(11.57\%\) \\
   \bottomrule
    \end{tabular}%
  }
\end{table}
\subsection{MV segment}
\label{sec:guidelines}

The admittance matrix of the MV line is:
\begin{equation}\label{model}
\begin{bmatrix}
I_a\\
I_b\\
I_c\\
\end{bmatrix}=
\begin{bmatrix}
Y_{aa} & Y_{ab} & Y_{ac} \\
Y_{ba} & Y_{bb} & Y_{bc} \\
Y_{ca} & Y_{cb} & Y_{cc} \\
\end{bmatrix}
\begin{bmatrix}
V1_a-V2_a \\
V1_b-V2_b\\
V1_c-V2_c\\
\end{bmatrix}
\end{equation}
The admittance matrix $Y_{abc}$ in (\ref{model}) can be diagonalized in two steps. First, apply the Fortescue transformation to transform $Y_{abc}$ to symmetrical coordinates:

\begin{equation}\label{line012}
\begin{aligned}
&I_{abc}=Y_{abc}\,V_{abc}\\
&T^{abc}_{012}\,I_{012}=Y_{abc}\,T^{abc}_{012}\,V_{012}\\
&I_{012}=T^{012}_{abc}\,Y_{abc}\,T_{012}^{abc}\,V_{012}\\
&\therefore Y_{012}=T^{012}_{abc}\,Y_{abc}\,T_{012}^{abc}
\end{aligned}
\end{equation}
where $T^{012}_{abc}$, $T^{abc}_{012}$ are the phase to Fortescue and Fortescue to phase transformation matrices of order three ($a_3=\frac{2 \pi}{3}$):
\begin{equation}\label{T01221}
\begin{aligned}
    &T^{012}_{abc}=
\frac{1}{3}
\begin{bmatrix}
1 & 1 & 1 \\
1 & a_3 & a_3^2 \\
1 & a_3^2 & a_3 \\
\end{bmatrix}&
    & T^{abc}_{012}=(T^{012}_{abc})^{-1}=
\begin{bmatrix}
1 & 1 & 1 \\
1 & a_3^2 & a_3 \\
1 & a_3 & a_3^2 \\
\end{bmatrix}
\end{aligned}
\end{equation}
Second, represent the mutual coupling in (\ref{model}) as current injections:
\begin{equation}\label{model012}
\begin{bmatrix}
I_0\\
I_1\\
I_2\\
\end{bmatrix}=
\begin{bmatrix}
Y_{00} & 0 & 0 \\
0 & Y_{11} & 0 \\
0 & 0 & Y_{22} \\
\end{bmatrix}
\begin{bmatrix}
V1_0-V2_0 \\
V1_1-V2_1\\
V1_2-V2_2\\
\end{bmatrix}
+
\begin{bmatrix}
\Delta I_0 \\
\Delta I_1\\
\Delta I_2\\
\end{bmatrix}
\end{equation}
where the current injections $\Delta I_{012}$ are:
\begin{equation}\label{injs}
\begin{aligned}
\Delta I_0 = Y_{01}(V1_0-V2_1)+Y_{02}(V1_0-V2_2)\\
\Delta I_1= Y_{10}(V1_1-V2_0)+Y_{12}(V1_1-V2_2)\\
\Delta I_2= Y_{20}(V1_2-V2_0)+Y_{21}(V1_2-V2_1)\\
\end{aligned}
\end{equation}
this procedure is only feasible under the following circumstances \cite{2}, \cite{3}, \cite{FAST}, \cite{mam1}, \cite{contributions}: (a) the magnitudes of $V_1$ and $I_1$ are significantly larger than those of the zero and negative sequences, (b) in symmetrical coordinates, the diagonal elements of the line admittance matrix are significantly larger than the off-diagonal ones, indicating a weak mutual coupling in the symmetrical coordinates.
\
\subsection{LV segment}
The bus admittance matrix of the LV segment is:
\begin{equation}\label{Tmodel}
\begin{bmatrix}
I3_{abcn} \\
I4_{abcn}\\
\end{bmatrix}=
\begin{bmatrix}
Y^{line}_{abcn}+ Y^{shunt}_{abcn}& -Y^{line}_{abcn}\\
-Y^{line}_{abcn}  & Y^{line}_{abcn}+Y^{shunt}_{abcn}\\
\end{bmatrix}
\begin{bmatrix}
V3_{abcn} \\
V4_{abcn}\\
\end{bmatrix}
\end{equation}
where $Y^{line}_{abcn}$ represents the series line admittance and $Y^{shunt}_{abcn}$ represents the shunt admittance, which consists of the grounding resistance and shunt capacitance. Typically, $Y^{shunt}_{abcn}$ is orders of magnitude smaller than $Y^{line}_{abcn}$. Consequently, it is more computationally efficient to diagonalize $Y^{line}_{abcn}$, whereas $Y^{shunt}_{abcn}$ can be modeled as current injections following (\ref{model012}). The transformation matrices that diagonalize $Y^{line}_{abcn}$ were calculated using the eigenvector decomposition presented in Sub-section \ref{diagss}. Here, $T_{eig}^{abcn}$ transforms the currents from eigen-basis to phase coordinates, $T_{abcn}^{eig}$ transforms the phase voltages to eigen-basis coordinates and ${v}_{i,j}$ is the $i_{th}$ element of the $j_{th}$ eigenvector. 
\begin{equation}
I_{abcn}=T_{eig}^{abcn} I_{eig}\\    
\end{equation}
\begin{equation}\label{Teigtoabcn}
\begin{bmatrix}
I_a \\
I_b \\
I_c \\
I_n
\end{bmatrix}
=
\begin{bmatrix}
{v}_{0,0} & {v}_{0,1} & {v}_{0,2} &  {v}_{0,4} \\
{v}_{1,0} & {v}_{1,1} & {v}_{1,2} &{v}_{1,4} \\
{v}_{2,0} & {v}_{2,1} & {v}_{2,2} &  {v}_{2,4} \\
{v}_{4, 1} & {v}_{4, 2} & {v}_{4, 3} & {v}_{4,4} 
\end{bmatrix}
\begin{bmatrix}
I_{\lambda 1} \\
I_{\lambda 2} \\
I_{\lambda 3} \\
I_{\lambda 4}
\end{bmatrix}
\end{equation}
\begin{equation}
V_{eig}=T_{abcn}^{eig} V_{abcn}\\    
\end{equation}
\begin{equation}\label{Tabcntoeig}
\begin{bmatrix}
V_{\lambda 1} \\
V_{\lambda 2} \\
V_{\lambda 3} \\
V_{\lambda 4}
\end{bmatrix}
=
\begin{bmatrix}
{v}_{0,0} & {v}_{0,1} & {v}_{0,2} &  {v}_{0,4} \\
{v}_{1,0} & {v}_{1,1} & {v}_{1,2} &{v}_{1,4} \\
{v}_{2,0} & {v}_{2,1} & {v}_{2,2} &  {v}_{2,4} \\
{v}_{4, 1} & {v}_{4, 2} & {v}_{4, 3} & v_{4,4} 
\end{bmatrix}^{-1}
\begin{bmatrix}
V_a \\
V_b \\
V_c \\
V_n
\end{bmatrix}
\end{equation}
\subsection{ MV/LV transformer}
The current injections of the MV segment to the three-phase four-wire LV segment can be accounted for by applying correct phase mapping \cite{15}. This method is adopted in the modelling of the MV/LV transformer. The bus admittance model of the delta/wye-grounded transformer is shown in (\ref{Tmodel}).  
\begin{equation}\label{Tmodel}
\begin{bmatrix}
I^{D_{1\times3}}_{abc} \\
I^{Y_{1\times4}}_{abcn}\\
\end{bmatrix}=
\begin{bmatrix}
Y^{3\times3}_{DD} & Y^{3\times4}_{DY}\\
Y^{4\times3}_{YD}  & Y^{4\times4}_{YY}\\
\end{bmatrix}
\begin{bmatrix}
V^{D_{1\times3}}_{abc} \\
V^{Y_{1\times4}}_{abcn}\\
\end{bmatrix}
\end{equation}
where the sub-matrices $Y_{DD}$, $Y_{DY}$, $Y_{YD}$, $Y_{YY}$ are shown in (\ref{TDYmodel}), $y_t$ is the leakage admittance, $Y_{gr}$ is the grounding admittance $Y_{YD}=Y^{T}_{DY}$ \cite{comp}: 
\begin{equation}\label{TDYmodel}
\resizebox{0.91\columnwidth}{!}{$
\begin{aligned}
    &Y_{DD} =
    \frac{1}{3}
    \begin{bmatrix}
    2y_{t} & -y_{t} & -y_{t} \\
    -y_{t} & 2y_{t} & -y_{t} \\
    -y_{t} & -y_{t} & 2y_{t} \\
    \end{bmatrix} , 
    Y_{DY} =
    \frac{1}{\sqrt{3}}
    \begin{bmatrix}
    -y_{t} & y_{t} & 0 & 0\\
    0 & -y_{t} & y_{t}& 0 \\
    y_{t} & 0 & -y_{t}& 0 \\
    \end{bmatrix}\\
    &Y_{YgYg} =
    \begin{bmatrix}
    y_{t} & 0 & 0 &-y_{t}\\
    0 & y_{t} & 0 &-y_{t}\\
    0 & 0 & y_{t} &-y_{t}\\
    -y_{t} &-y_{t} &-y_{t}&3y_{t}+Y_{gr}\\
    \end{bmatrix}
\end{aligned}
$}
\end{equation}
\begin{figure}[t!] 
\centering
\includegraphics[ width=3.3 in]{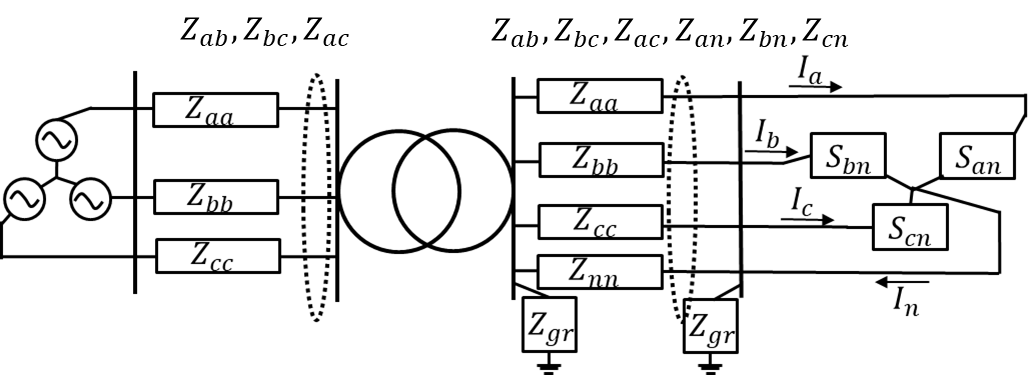}
\caption{ Integrated MV/LV 4-bus feeder}
\label{fig:MV_LV}
\end{figure}\
 The transformation of the sub-matrices from phase coordinates to symmetrical and eigen-basis coordinates is performed as follows:
\begin{description}
\item[1)] The self admittance matrix of the primary side $Y^{{abc}_{3\times3}}_{DD}$ is transformed to symmetrical coordinates following (\ref{line012})

\item[2)] The mutual admittance matrix $Y^{{abc}_{3\times3}}_{DYg}$, obtained by eliminating the fourth column from the original matrix in (\ref{TDYmodel}), and relates the primary side current $I^{D_{1\times3}}_{abc}$ to the secondary side phase voltage $V^{Yg_{1\times3}}_{abc}$, is transformed to symmetrical and eigen-basis coordinates as follows:

\begin{equation}\label{TDY2}
\begin{aligned}
&I^{D_{1\times3}}_{abc}=Y^{{abc}_{3\times3}}_{DYg}\,V^{Yg_{1\times3}}_{abc}\\
&T_{012}^{abc}\,I^{D_{1\times3}}_{012}=Y^{{abc}_{3\times3}}_{DYg}\,T_{eig}^{abc}\,V^{Yg_{1\times4}}_{eig}\\
&I^{D_{1\times3}}_{012}=T^{012}_{{abc}}\,Y^{{abc}_{3\times3}}_{DYg}\,T_{eig}^{abc}\,V^{Yg_{1\times4}}_{eig} \\
&\therefore Y^{{012/eig}_{3\times4}}_{DYg}=T^{012}_{{abc}}\,Y^{{abc}_{3\times3}}_{DYg}\,T_{eig}^{abc} \\
\end{aligned}
\end{equation}

where the matrix $T^{abc}_{{eig}}$ maps the LV side phase voltages to eigen-basis coordinates. The $T^{abc}_{eig}$ is found by removing the fourth row of the matrix $T^{abcn}_{eig}$ shown in (\ref{Teigtoabcn}), which corresponds to the neutral components. 

\item[3)] The mutual admittance matrix $Y^{{abc}_{3\times3}}_{YD}$, obtained by eliminating the fourth row and relates the secondary side current $I^{Yg_{1\times3}}_{abc}$ to the primary side phase voltage $V^{D_{1\times3}}_{abc}$, is transformed to symmetrical and eigen-basis coordinates as follows:

\begin{equation}\label{TDY3}
\begin{aligned}
&I^{Yg_{1\times3}}_{abc}=Y^{{abc}_{3\times3}}_{YgD}\,V^{D_{1\times3}}_{abc}\\
&I^{Yg_{1\times3}}_{abc}=Y^{{abcn}_{3\times3}}_{YgD}\,T_{012}^{abc}\,V^{D_{1\times3}}_{012}\\
&T^{eig}_{{abc}}\,I^{Yg_{1\times3}}_{abc}=T^{eig}_{{abc}}\,Y^{{abc}_{3\times3}}_{YgD}\,T_{012}^{abc}\,V^{D_{1\times3}}_{012} \\
&I^{Yg_{1\times4}}_{eig}=T^{eig}_{{abc}}\,Y^{{abc}_{3\times3}}_{YgD}\,T_{012}^{abc}\,V^{D_{1\times3}}_{012} \\
&\therefore Y^{{eig/012}_{4\times3}}_{YgD}=T^{eig}_{{abc}}\,Y^{{abc}_{3\times3}}_{YgD}\,T_{012}^{abc} \\
\end{aligned}
\end{equation}
where the matrix $T^{eig}_{{abc}}$ maps the LV side phase currents  to eigen-basis coordinates. The $T^{eig}_{{abc}}$ is found by removing the fourth column of the matrix $T^{eig}_{abcn}$ shown in (\ref{Tabcntoeig}), which corresponds to the neutral components.

\item[4)] The self admittance matrix of the secondary side $Y^{4\times4}_{YgYg}$ is transformed to eigen-basis coordinates by transforming the secondary side voltages $V^{1\times4}_{abcn}$ and currents $I^{1\times4}_{abcn}$ to eigen-basis coordinates as follows:
\begin{equation}\label{TDY4}
\begin{aligned}
&I^{Yg_{1\times4}}_{abcn}=Y^{{abcn}_{4\times4}}_{YgYg}\,V^{Yg_{1\times4}}_{abcn}\\
&T^{abcn}_{{eig}}\,I^{Yg_{1\times4}}_{eig}=Y^{{abcn}_{4\times4}}_{YgYg}\,T_{eig}^{abcn}\,V^{Yg_{1\times4}}_{eig}\\
&I^{Yg_{1\times4}}_{eig}=T^{eig}_{{abcn}}\,Y^{{abc}_{4\times4}}_{YgYg}\,T_{eig}^{abc}\,V^{D_{1\times4}}_{eig} \\
&\therefore Y^{{eig}_{4\times4}}_{YgYg}=T^{eig}_{{abcn}}\,Y^{{abc}_{4\times4}}_{YgYg}\,T_{eig}^{abcn} \\
\end{aligned}
\end{equation}
where $T^{eig}_{{abcn}}$ equals the inverse of $T^{abcn}_{{eig}}$.
    \end{description} 
    
The bus admittance matrix in symmetrical and eigen-basis coordinates ($Y^{012/eig}_{BUS}$) of the MV/LV feeder in Fig. \ref{fig:MV_LV} is shown in (\ref{MYBus}).
\begin{equation}
\resizebox{0.9\columnwidth}{!}{$
\begin{bmatrix}
I^{MV}_{012} \\
I^{D}_{012} \\
I^{Yg}_{eig} \\
I^{LV}_{eig} \\
\end{bmatrix}=
\begin{bmatrix}
Y^{012}_{MV} & -Y^{012}_{MV} & 0 & 0 \\
-Y^{012}_{MV} & Y^{012}_{MV} + Y^{012}_{DD} & Y^{eig}_{DYg} & 0 \\
0 & Y^{eig}_{DYg} & Y^{eig}_{YgYg} + Y^{eig}_{LV} & -Y^{eig}_{LV} \\
0 & 0 & -Y^{eig}_{LV} & Y^{eig}_{LV} \\
\end{bmatrix}
\begin{bmatrix}
V^{MV}_{012} \\
V^{D}_{012} \\
V^{Yg}_{eig} \\
V^{LV}_{eig} \\
\end{bmatrix}
$}
\label{MYBus}
\end{equation}
\section{Power Flow}
Power flow calculations are performed using different methods. In this study, the Y\textsubscript{BUS} method was selected \cite{Das}. 
\subsection{ Y\textsubscript{BUS} method}
The main steps of the Y\textsubscript{BUS} method are:
\begin{itemize}
    \item[1)] Build the system bus admittance matrix in symmetrical and eigen-basis coordinates $Y^{012/eig}_{BUS}$ following the approach in Section \ref{bui}. Set the power flow iteration counter ($C$) to 1.
    \item[2)]   Perform LU factorization with approximate minimum degree permutation to decompose $Y^{012/eig}_{BUS}$ into  $Y^{012/eig}_{L}$,  $Y^{012/eig}_{U}$ \cite{AMDLU}.      
    \item[3)] Assume initial voltage magnitudes of 1.0 pu for phases and 0 for the neutral, and initial phase angles of $0^\circ$, $-120^\circ$, $-240^\circ$, $0^\circ$ for phase a, b, c and the neutral, respectively.
    \item[4)]  Calculate the ZIP load model current injections ($I^{inj}_{ph}$), and transform the currents to symmetrical and eigen-basis coordinates.
\begin{equation}\label{initialV}
\begin{aligned}
&I^{inj}_{ph}=-(\frac{V_{ph}}{Z^{load}_{ph}}+I^{load}_{ph}+\frac{S^*_{ph}}{V^*_{ph}})\\
&I^{inj}_n=-(I^{inj}_{a}+I^{inj}_{b}+I^{inj}_{c})\\
\end{aligned}
\end{equation}
    \item[5)]Solve for the voltages in (\ref{syst}) using forward/backward substitution.
    \begin{equation}\label{syst}
\begin{bmatrix}
    I_{012}\\I_{eig}  \end{bmatrix}  =
    [Y^{012/eig}_{L}]
  [Y^{012/eig}_{U}]
     \begin{bmatrix}V_{012}\\V_{eig}\end{bmatrix}
   \end{equation}
    \item[6)]Transform the calculated voltages to phase coordinates, calculate the voltage error:
\begin{equation}\label{initialV}
\begin{aligned}
&V_{er}(C)=|V_{ph}(C-1)-V_{ph}(C)|
\end{aligned}
\end{equation}
    \item[7)] 
    If the maximum voltage error falls under a defined threshold ($max(V_{er}(C))<=\epsilon$) terminate the process, else increase C by 1 go to step 4. 
\end{itemize} 
\subsection{ Decomposed Y\textsubscript{BUS} method} \label{Dec}
\begin{figure}[t!] 
\centering
\includegraphics[width=3 in]{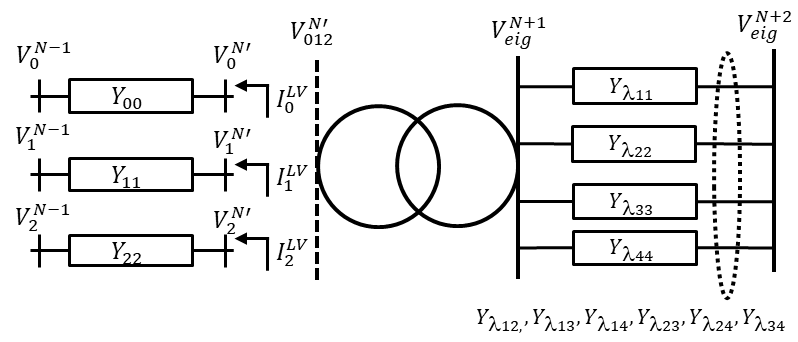}
\caption{ Decomposed network in symmetrical and eigen-basis coordinates.}
\label{MVdecoupled}
\end{figure}

\begin{figure}[t!] 
\centering
\includegraphics[width=3.2 in]{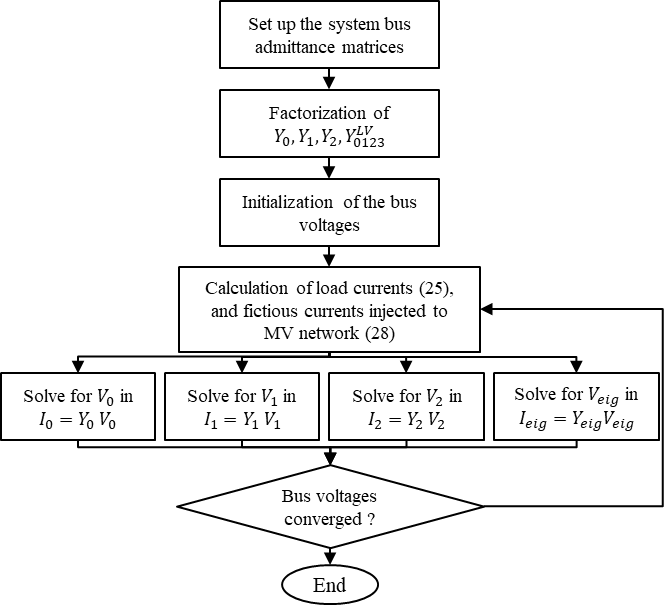}
\caption{Flow chart of the decomposed Y\textsubscript{BUS} method.}
\label{process}
\end{figure}

The Y\textsubscript{BUS} method can be made more efficient by decomposing the integrated MV/LV distribution network to smaller sub networks that can be solved in parallel \cite{mam2}. Fig. \ref{MVdecoupled} shows the decomposition of the MV/LV network. The symmetrical MV network is decoupled into positive, negative and zero sequence networks. The LV networks are represented by equivalent current injections ($I^{LV}_{012}$). The power flow solution of the decoupled MV networks determine the value of the fictitious slack bus ($V^{N'}_{eig}$) of the decomposed LV network. Also, the power flow solution of the decomposed LV network determines the equivalent current injections ($I^{LV}_{012}$) to the decoupled MV networks, as shown in (\ref{I012Inj}). It is important to note that the mutual coupling in the LV segment arises from $Y^{shunt}$, which appears in the diagonal blocks of the Y\textsubscript{BUS} matrix. Fig. \ref{process} shows the decomposed Y\textsubscript{BUS} method. 
\begin{equation}\label{I012Inj}
I^{LV}_{012}=[Y^{DD}_{012} \quad Y^{{012/eig}}_{DYg}] \begin{bmatrix}V_{012}\\V_{eig} \end{bmatrix}
\end{equation}
\section{Case Studies}
\begin{figure}[t!]  
  \centering
  \begin{subfigure}[b]{0.49\textwidth}
    \centering
    \includegraphics[width=3.4in]{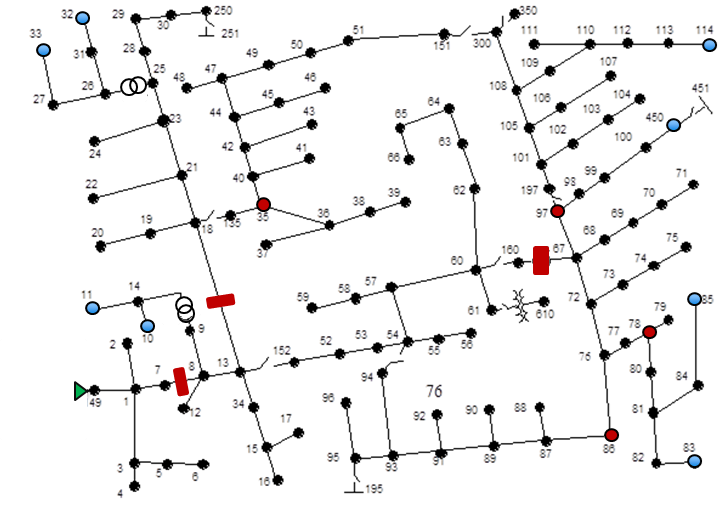}
    \caption{}
    \label{fig:square1}
  \end{subfigure}
  \hspace{0.05\textwidth}
  \begin{subfigure}[b]{0.49\textwidth}
    \centering
    \includegraphics[width=3.4in]{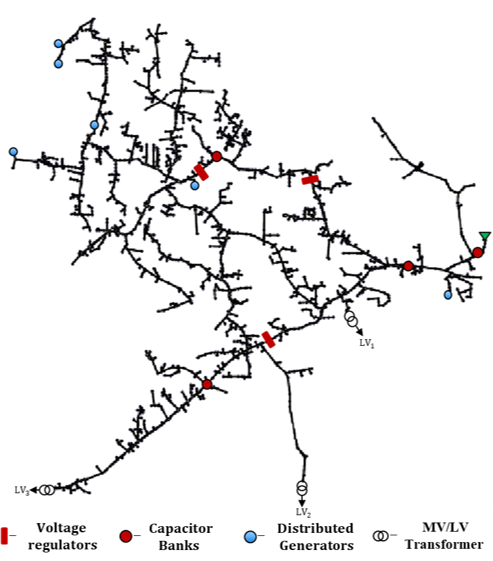}
\caption{}
    \label{fig:square2}
  \end{subfigure}
  \vspace{0.1cm}
  \begin{subfigure}[b]{0.49\textwidth}
    \centering
    \includegraphics[width=3in]{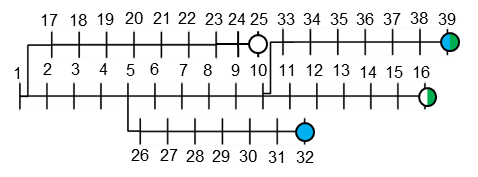}
    \caption{}
    \label{fig:rectangle}
  \end{subfigure}
  \caption{Modified (a) IEEE-123 bus, (b) IEEE-8500 node network, (c) Configuration of the three unsymmetrical LV feeders connected to IEEE-8500 node network. The distributed generators of the first LV feeder are represented by green circles, the second LV feeder by blue circles, and the third LV feeder by white circles.}
  \label{netow}
\end{figure}

In this section, the accuracy and memory requirements of the proposed power flow method are assessed with reference to phase coordinates power flow methods presented in  \cite{ALAM} and \cite{Das}. The performance of the proposed power flow method was evaluated using IEEE-4 bus feeder and modified versions of the IEEE-123 bus and IEEE-8500 node network. The modifications are:  (a) the main feeder starting from the slack bus is assumed to be transposed with a three-wire configuration. (b) LV three-phase four-wire multi-grounded feeders are connected to the main feeder by delta/wye-grounded transformers at different locations as shown in Fig. \ref{netow}. The LV feeder and delta/wye-grounded transformers data of the IEEE-4 bus feeder are adopted to model the LV segments, and the grounding resistance is assumed to be 100 $\Omega$. The power flow methods were programmed in Matlab R2021b running on a Lenovo laptop having a 2.7 GHz, 6 cores Intel Core i7 processor with 12 MB L3 cache and 16 GB of RAM.
\subsection{Accuracy} \label{acc}
Fig. \ref{fig:MV_LV} shows the IEEE-4 bus feeder studied in \cite{ALAM}, which has a grounding impedance ($Z_{gr}=0.3\hspace{0.1cm} \Omega$). This feeder is used to test the accuracy of the proposed power flow method, and showcase the limitations of the Kron reduced model. The results of the proposed decoupled Y\textsubscript{BUS} and Kron reduced Y\textsubscript{BUS} power flow methods are compared to the phase coordinates Newton current injection method (NCIM) presented in \cite{ALAM}. The threshold $\epsilon$ of the power flow calculation was $10^{-6}$. Table \ref{tab:my-table} shows the voltage magnitudes and phase angles of the network buses. The result of the decomposed Y\textsubscript{BUS} method shows a slight error ($10^{-4}$ pu). This error is a result of the use of fictitious slack buses and current injection in the decomposed MV/LV network \cite{mam2}. In the Kron reduced Y\textsubscript{BUS} method the neutral node is eliminated, and the neutral-to-ground voltage is neglected in the calculation of the load current ($I_{Load_{KR}}$) as shown in (\ref{Iload}). Consequently, this method introduces significant errors in the voltage magnitude ($10^{-2}$ pu). 
\begin{equation}\label{Iload}
\begin{aligned}
&I_{Load}=-(\frac{S^*_{ph}}{(V_{ph}-V_{n})^*})\\
&I_{Load_{KR}}=-(\frac{S^*_{ph}}{V^*_{ph}})\\
\end{aligned}
\end{equation}

It is important to note that the relatively low voltages at Bus 4 are due to the fact that the IEEE 4-bus test feeder is heavily loaded. Nevertheless, the obtained per unit values are consistent with the results reported in \cite{ALAM}.
\subsection{Reduction of memory requirement}
The memory requirement of the power flow calculation is the number of bytes required to store the non-zero elements of the lower and upper triangular matrices, the transformation matrices to symmetrical and eigen-basis coordinates. Each non-zero element is represented by a double precision complex number, necessitating 16 bytes for storage. The percentage of memory reduction achieved by transforming the phase coordinates to symmetrical and eigen-basis coordinates is presented in Table \ref{mem}. The number of non-zero elements in the symmetrical and eigen-basis coordinates is equal to the sum of the non-zero elements of the transformation matrices, the three decoupled symmetrical networks, and the LV networks. 
\section{Volt/Var Control} \label{VVC}
First, let us consider the DCD algorithm for a VVC problem. The objective function of the DCD is shown in (\ref{DCD}):
\begin{equation}\label{DCD}
\begin{aligned}
f(x) &= \sum_{i=1}^{n} \sum_{j=1}^{n} V(x)_i^{re} Y_{ij}^{re} V(x)_j^{re} \\
&\quad + \sum_{i=1}^{n} \sum_{j=1}^{n} V(x)_i^{im} Y_{ij}^{re} V(x)_j^{im} \\
&\quad + C_V \sum_{i=1}^{n} \max \left(0, V(x)_i - V_i^{max} \right) \\
&\quad + C_V \sum_{i=1}^{n} \max \left(0, V_i^{min} - V(x)_i \right) \\
&\quad + C_I \sum_{ij \in BR} \max \left(0, I(x)_{ij} - I_{ij}^{max} \right)
\end{aligned}
\end{equation}
where $V(x)_i$ is the magnitude of the bus voltage at node $i$,  $V(x)^{re}$ and $V(x)^{im}$ are the real and imaginary parts of the bus voltage, $Y_{ij}^{re}$ is the real part of the bus admittance matrix element between nodes $i$ and $j$, $I(x)_{ij}$ is the magnitude of the line current, $C_V$ and $C_I$ are penalty coefficients for voltage and line current magnitudes violations, and $x$ represents the control variables, which include transformer tap settings, reactive power of distributed generators, and shunt capacitors.

The sensitivity-based DCD algorithm is an improved version of the DCD algorithm, in which the sensitivity of the objective function to each control variable ($ \frac{\partial f}{\partial x}$) is calculated analytically instead of approximating it through numerical perturbation. This reduces the number of required power flow calculations by half. A method to calculate the sensitivity factors of the objective function for the positive sequence model of the network is presented in \cite{sensDCD}. This method was applied to multi-phase models of the network and control variables in \cite{three-phase}. The sensitivity-based DCD algorithm is summarized in the following steps:
\begin{table}[t]
  \centering
  \caption{\textsc{Bus Voltages and Errors for The IEEE-4 Bus Feeder}}
   \renewcommand{\arraystretch}{2}
  \label{tab:my-table}
  \resizebox{\linewidth}{!}{%
  \Large 
  \fontsize{100}{100}\selectfont 
    \begin{tabular}{ccccc}
      \toprule
        & Phase &   Bus 2 & Bus 3 & Bus 4 \\
      \midrule
      \multirow{4}{*}{NCIM \cite{ALAM} }  & a &$0.9902\angle -0.244^\circ$ &  $0.9668\angle -33.769^\circ$ & $0.8972\angle -35.023^\circ$ \\
                                & b &  $0.9883\angle -120.548^\circ$ &  $0.9623\angle -152.239^\circ$ & $0.8354\angle -156.254^\circ$ \\ 
                                & c  & $0.9870\angle 119.756^\circ$ &  $0.8813\angle 84.712^\circ$ & $0.7235\angle 72.185^\circ$ \\
                                & n  & - &  $0.0298\angle 86.276^\circ$ & $0.0298\angle -93.724^\circ$ \\
     \midrule 
 & a &$0.9902\angle -0.244^\circ$ &  $0.9668\angle -33.769^\circ$ & $0.8972\angle -35.023^\circ$ \\
                 {Decomposed }          & b &  $0.9884\angle -120.548^\circ$ &  $0.9623\angle -152.239^\circ$ & $0.8354\angle -156.254^\circ$ \\ 
     { $Y^{012/eig}_{BUS}$}    & c  & $0.9870\angle 119.755^\circ$ &  $0.8813\angle 84.713^\circ$ & $0.7236\angle 72.188^\circ$ \\
                                & n  & - &  $0.0298\angle 86.290^\circ$ & $0.0298\angle -93.710^\circ$ \\
                                 \midrule
  & a &$0.9898\angle -0.274^\circ$ &  $0.9536\angle -32.398^\circ$ & $0.8980\angle -34.245^\circ$ \\
                  {Kron Reduced}          & b &  $0.9888\angle -120.488^\circ$ &  $0.9416\angle -153.814^\circ$ & $0.8061\angle -157.035^\circ$ \\ 
      {  $Y^{012}_{BUS}$}    & c  & $0.9882\angle 119.722^\circ$ &  $0.9218\angle 85.177^\circ$ & $0.7700\angle 73.394^\circ$ \\
                                & n  & - &  - &  - \\
                                 \midrule
 \multirow{2}{*}{Error of}  & a &$0.0000 \angle 0.000^\circ$ &  $0.0000 \angle                                            0.000^\circ$ & $0.0000 \angle 0.000^\circ$ \\
             {Decomposed }                   & b &  $0.0001 \angle 0.000^\circ$ &  $0.0000 \angle 0.000^\circ$ & $0.0000 \angle 0.000^\circ$ \\ 
              { $Y^{012/eig}_{BUS}$}                 & c  & $0.0000 \angle -0.001^\circ$ &  $0.0000 \angle -0.001^\circ$ & $0.0001 \angle 0.003^\circ$ \\
                                & n  &- &  $0.0000 \angle 0.014^\circ$ & $0.0000 \angle 0.014^\circ$ \\
                                 \midrule
  \multirow{2}{*}{Error of}  & a &$0.0004 \angle 0.030^\circ$ &  $0.0132 \angle 1.371^\circ$ & $0.0012 \angle 0.778^\circ$ \\
                       {Kron Reduced}          & b &  $0.0018 \angle 0.060^\circ$ &  $0.0187 \angle 1.575^\circ$ & $0.0006 \angle 0.781^\circ$ \\ 
                    {$Y^{012}_{BUS}$}          & c  & $0.0008 \angle 0.034^\circ$ &  $0.0595 \angle 0.465^\circ$ & $0.0009 \angle 1.209^\circ$ \\
                             & n  & - &  - &  - \\

      \bottomrule
    \end{tabular}%
  }
\end{table}
\begin{table}[t]
  \centering
  \Large
  \caption{\textsc{Memory Requirements}}
    \renewcommand{\arraystretch}{2}
  \resizebox{1\linewidth}{!}{%
 \fontsize{30}{30}\selectfont 
    \begin{tabular}{cccccc}
      \toprule
      \multirow {2}{*}{Network} &\multirow {2}{*}{Method} & \multirow {1}{*}{non-zeros} \multirow {1}{*}& {non-zeros} & \multirow {1}{*}{Total Number} &\multirow {1}{*}{Percentage of}  \\
                 &         &       {in L}              &         {in U}          &   {of non-zeros}       & {Reduction}              \\
      \midrule
    \multirow {4}{*}{IEEE-123}  & $Y^{abcn}_{BUS}$  \cite{Das}& 1805 & 2301 &   4106 & \multirow {4}{*}{56.66\%} \\
                                    & \multirow{3}{*}{$Y^{012/eig}_{BUS}$}  & \multirow {1}{*}{$L_{0,1,2}=219$} & \multirow {1}{*}{$U_{0,1,2}=246$} &    \multirow{3}{*}{1781} &  \\ 
                                    & & $L_{LV1}=66$& $U_{LV1}=70$ & & \\
                                    & & $L_{LV2}=106$& $U_{LV2}=124$ & & \\
    \bottomrule
        \multirow {3}{*}{IEEE-8500}  & $Y^{abcn}_{BUS}$  \cite{Das}& 39391 & 53588 &   92979 &  \multirow {3}{*}{59.53\%}  \\
                            &  \multirow{2}{*}{$Y^{012/eig}_{BUS}$}  & \multirow {1}{*}{$L_{0,1,2}=4989$} & \multirow {1}{*}{$U_{0,1,2}=5797$} &    \multirow{2}{*}{37629} &  \\ 
                                    & & $L_{LV_{1,2,3}}=768$& $U_{LV_{1,2,3}}=983$ & & \\ 
    \bottomrule
    \end{tabular}%
  }
 \label{mem}
\end{table}
\begin{itemize}
\item\textbf{Step 1:}  
Initialize the control variables and set the iteration counter to 1.

\item\textbf{Step 2:}  
Form the bus admittance matrix accounting for the initial settings of the control variables.

\item\textbf{Step 3:}  
Compute the base-case power flow using the decomposed Y\textsubscript{BUS} method presented in Sub-section \ref{Dec}. Then calculate the objective function using (\ref{DCD}) and assign its value to the Best Objective Function (BOF).
\item\textbf{Step 4:}
Determine the initial Descent Directions, indexed by \( \text{DD} = 1, \ldots, \text{NDD} \). Calculate the sensitivity of the objective function $\left( \frac{\partial f}{\partial x} \right)$ with respect to each control variable. Initialize \( \text{DD} = 1 \) and set the index of the Best Descent Direction to zero: \( \text{BDD} = 0 \). 
\item\textbf{Step 5:}
For each control variable with a positive sensitivity factor, reduce its value by one step, provided it has not reached its lower bound. For control variables with a negative sensitivity factor, increase its value by one step, if it has not hit its upper limit.
\item\textbf{Step 6:}  
 Apply the compensation technique presented in \cite{injs} to account for the change in the control variable.
\item\textbf{Step 7:}
Perform the power flow calculation using the decomposed Y\textsubscript{BUS} method and calculate the Iteration Objective Function (IOF). Restore the control variable to its initial value.
\item\textbf{Step 8:}  
If \( \text{IOF} < \text{BOF} \), then set \( \text{BOF} = \text{IOF} \) and \( \text{BDD} = \text{DD} \); otherwise, keep them at their current values.
\item\textbf{Step 9:}  
If \( \text{BD} < \text{NBD} \), increment \( \text{BD} \) by 1 and return to Step 5; otherwise, proceed to Step 10.
\item\textbf{Step 10:}  
If \( \text{BDD} = 0 \), terminate the search (indicating that the optimization is complete); otherwise, adjust the control variable according to the direction indicated by \( \text{BDD} \), increment the DCD iteration counter by 1, and return to Step 4.
\end{itemize}

\subsection{Proposed power flow for VVC application}
VVC using the sensitivity-based DCD algorithm is adopted to demonstrate the computational advantage of the proposed power flow methods in a practical scenario. The locations of the networks control devices are illustrated in Fig. \ref{netow}. The number of taps per transformer is typically 33, the capacity of a switched capacitor is chosen to be 200 kVAr with a step length of 50 kVAr in the modified IEEE-8500 network, and 180 kVAr with a step length of 20 kVAr in the modified IEEE-123 network, and the resolution for DG VAr output is $10^{-3}$ pu \cite{sensDCD}.

The computational advantage of the proposed power flow methods is presented in Table \ref{times}, which shows the number of DCD iterations, power flow runs and the computation time with the speed up factor (SUF). The computation time is the time needed for: (a) the factorization of the Y\textsubscript{BUS} matrix, (b) the calculation of $I=Y_{L}Y_{U} V$ using forward/backward substitution, (c) the conversion of voltages and currents to phase and symmetrical and eigen-basis coordinates respectively \cite{1}. It is important to note that Y\textsubscript{BUS} is factorized only once, as any alterations in the tap positions and capacitor banks are accounted for using the compensation technique \cite{injs}. 

Compared to the $Y^{abcn}_{BUS}$ method, both the full and decomposed $Y^{012/eig}_{BUS}$ methods yield computational improvements in calculating the power flows for the IEEE-123 and IEEE-8500 systems. The decomposed $Y^{012/eig}_{BUS}$ method achieved speed-up factors of 2.78 and 3.63 for the IEEE-123 and IEEE-8500 systems, respectively. The full $Y^{012/eig}_{BUS}$ method achieved speed-up factors of 1.38 and 1.56, relative to the $Y^{abcn}_{BUS}$ method. These computational improvements are due to fewer non-zero fill-ins when building the lower and upper triangular matrices in symmetrical and eigen-basis coordinates. For example, in the IEEE-8500 test system, the $Y^{abcn}_{BUS}$ method results in 39,391 non-zero elements in the lower triangular matrix $L_{abc}$ and 53,588 in the upper triangular matrix $U_{abc}$. In the proposed coordinates, the factorization of the $Y^{012/eig}_{BUS}$ results in 17,643 and 20,614 non-zero elements in $L_{012/eig}$ and $U_{012/eig}$, respectively. The decomposed $Y^{012/eig}_{BUS}$ method further enhances the efficiency by enabling parallel computation. In this approach, six smaller networks are solved in parallel, with the largest systems having 4,989 non-zero elements in $L_{0,1,2}$ and 5,797 in $U_{0,1,2}$, as shown in Table \ref{mem}. Also, this parallelization significantly reduced the factorization time, from 7.84 ms for the $Y^{abcn}_{BUS}$ method to 0.88 ms for the decomposed $Y^{012/eig}_{BUS}$ method on the IEEE-8500 system. This result demonstrates the computational efficiency of the proposed methods in applications that require frequent refactorization. 

It is important to note that the parallel execution time is the maximum among the decomposed network execution times, reflecting the synchronization requirements of parallel computations. This approach does not account for parallel overheads, which may arise from task coordination and data allocation. However, the significant performance difference between the decomposed $Y^{012/eig}_{BUS}$ and $Y^{abcn}_{BUS}$ methods supports reasonable conclusions about the efficiency of the proposed methods. The number of DCD iterations differs slightly due to minor errors in the solution of the decomposed $Y^{012/eig}_{BUS}$. The maximum error in the voltage magnitude is less than $1 \times 10^{-5}$ pu at the secondary side of the step-down transformer of $LV_2$ in the IEEE-8500 and less than $4\times10^{-4} $ pu at Bus 10 in the IEEE-123, these errors were previously discussed in Sub-section \ref{acc}. The VVC solution of the decomposed  $Y^{012/eig}_{BUS}$ method differs only in the LV DG VAr outputs. In each DCD iteration the DG VAr output is adjusted by one step. The power flow solution of the full $Y^{012/eig}_{BUS}$ matches the solution of the phase coordinate model $Y^{abcn}_{BUS}$.

\begin{table}[t]
  \centering
  \Large
  \caption{\textsc{Number of DCD Iterations, Power Flow Runs, and Total Computational Time}}
\renewcommand{\arraystretch}{1.5} 
  \label{times}
  
  \resizebox{1.0\linewidth}{!}{
 \fontsize{20}{20}\selectfont 
    \begin{tabular}{cccccc}
      \toprule
      \multirow {3}{*}{Network} & \multirow {3}{*}{Method} & \multirow{1}{*}{DCD } & \multirow{1}{*}{Number of} & \multirow{1}{*}{Power flow} & \multirow{3}{*}{SUF }  \\
      & & {iterations} &{ power flow} &{executions} &   \\
      & & & {runs} & {time (s)}&  \\
      \midrule
     \multirow {2}{*}{IEEE-123}  & $Y^{abcn}_{BUS}$  \cite{Das} & 94 & 4325& 0.1743 &   
     \\& \( Y^{012/eig}_{BUS} \)   & 94 & 4325 &0.1261 &  1.38
     
     \\  & \( Y^{012/eig}_{BUS} \) (dec.)   & 96 & 4417 &0.0628 &  2.78 \\
   \bottomrule
        \multirow {3}{*}{IEEE-8500}  & $Y^{abcn}_{BUS}$ \cite{Das} & 51 & 2933 & 5.2599 &   \\
                                     &\( Y^{012/eig}_{BUS} \)  & 51 &  2933 & 3.3789 &  1.56\\
                                     &\( Y^{012/eig}_{BUS} \) (dec.)   & 52 &  2990 & 1.4476 &    3.63\\
    \bottomrule
    \end{tabular}%
  }
\end{table}
\section{Conclusion}
The proposed symmetrical and eigen-basis coordinates model for integrated three-wire MV and four-wire LV networks provides a significant improvement in power flow calculations. By utilizing eigenvector decomposition to diagonalize the admittance matrix of the LV segments, the model achieves substantial computational efficiency. The eigen-basis coordinate model should also replace the symmetrical coordinate model when the three-wire segment in the network is not symmetrical. The use of this model in VVC has demonstrated a significant reduction in the processing time and memory requirements of the sensitivity-based DCD algorithm. The decomposed method outperformed the phase coordinates method, being 2.78 to 3.63 times faster in performing DCD and reducing memory requirements by more than 50\%. This efficiency gain is due to the reduction in the number of non-zero elements and the effective parallel decomposition of the bus admittance matrix. These advantages make the proposed method well-suited for real-time power system operation and for training artificial intelligence models in VVC applications \cite{AI2}, \cite{AI3}.
\begingroup
\setlength{\itemsep}{0pt} 
\setlength{\parskip}{0pt} 
\renewcommand{\baselinestretch}{1}
\bibliographystyle{IEEEtran}
\bibliography{IEEEabrv,references}
\endgroup
\end{document}